\documentclass[12pt]{article}

\usepackage{subeqnarray}

\usepackage{amssymb}
\usepackage{stmaryrd}

\usepackage{amsfonts}
\usepackage{amsmath}
\def\half{{\textstyle\frac{1}{2}}}
\def\quart{{\textstyle\frac{1}{4}}}

\begin{document}

\title{Galilean Superconformal Symmetries}

\author{J.A. de Azc\'{a}rraga \\
Dept. of Theoretical Physics, University of Valencia, \\
46100 - Burjassot (Valencia) Spain\\
e-mail:azcarrag@ific.uv.es
\\ \\
J. Lukierski\\
Institute for Theoretical Physics, \\
University of Wroc{\l}aw, 50-205 Wroc{\l}aw, Poland\\
e-mail:lukier@ift.uni.wroc.pl}

\date{}
\maketitle

\begin{abstract}
We consider the non-relativistic  $c\to \infty$ contraction limit of the $(N=2k)$-extended $D=4$ superconformal algebra $\emph{su}(2,2;N)$, introducing in this way the non-relativistic $(N=2k)$-extended Galilean superconformal algebra. Such a Galilean superconformal algebra has the same number of generators as $\emph{su}(2,2|2k)$. The $\emph{usp}(2k)$ algebra describes  the non-relativistic internal symmetries, and the generators from the coset $\frac{\emph{u}(2k)}{\emph{usp}(2k)}$ become central charges after contraction.
 \end{abstract}

\section{Introduction}\label{secint}

There are two ways of enlarging the standard Galilei algebra by additional generators related with the conformal symmetries (see e.g. \cite{azcluk1})

i) One can add to the Galilean symmetries the dilatations and the one-parameter conformal transformations which determine the invariance group of free Schr\"{o}dinger  equation \cite{azcluk2}--\cite{azcluk6}. In such a way we obtain the so-called  Schr\"{o}dinger  algebra, which is the Galilean algebra enlarged by two generators $D$ (dilatations) and $K$ (time expansions). Recently, some authors have referred to the Schr\"{o}dinger symmetry as  Galilean conformal symmetry (see e.g. \cite{azcluk2,azcluk7}), but this approach does not lead to the typical features of conformal systems such as  vanishing masses, the presence of conformal space translations etc.

ii) Following the contraction of the Poincar\'{e}  algebra to the Galilei one, there was  performed as well  the non-relativistic contraction of relativistic conformal algebra (in $D$ dimensions $\emph{o}(D,2)$ to Galilean conformal algebra  \cite{azcluk8,azcluk9}. Denoting by $P_\mu = (P_0, P_i)$, $M _{\mu\nu}=(M_{ij}, M_{i0})$ ($\mu,\nu =0,1\dots D-1; \, i,j=1,2 \ldots d = D-1$) the Poincar\'{e} generators, the relativistic conformal algebra  also includes the generators of dilatations $D$ and those of the special conformal transformations $K_\mu =(K_0, K_i)$. If we rescale the relativistic generators in the following way \cite{azcluk8}\footnote{In comparison with \cite{azcluk8} we denote here the Galilean boosts by $B_i$, and the relativistic conformal translations generators by $K_\mu$.}

\begin{eqnarray}\label{azcluke1}
P_0 = \frac{H}{c} \, , &\qquad & M_{i 0} = cB_i\, ,
\cr
K_0 = cK \, , & \quad &  K_i = c^2 F_i \, ,
\end{eqnarray}
($P_i$, $M_{ij}$ and $D$ remaining unchanged) we obtain,  after performing the non-relativistic contraction limit $c\to \infty$, the following $\half (D+1)(D+2)$-dimensional Galilean conformal algebra
 \cite{azcluk8}--\cite{azcluk10}\footnote{It should be added that a family of non-relativistic conformal algebras was introduced in \cite{azcluk11}, with one member providing the relations~(\ref{azcluke2})--(\ref{azcluke6})}

\begin{equation}
\label{azcluke2}
 [H,P_i]=0 \, ,  \quad   [H,B_i]= P_i \, , \quad  [H,F_i]=2B_i\, ,
\end{equation}

\begin{equation}
[K, P_i]=-2B_i \, , \quad  [K,B_i]=F_i\, , \quad  [K, F_i]=0 \, ,
\label{azcluke3}
\end{equation}

\begin{equation}
[D,P_i]=- P_i \, , \quad  [D,B_i]=0 \, , \quad   [D,F_i]=F_i \, ,
\label{azcluke4}
\end{equation}
and

\begin{equation}\label{azcluke5}
[D,H] = - H \, , \quad [K,H]=-2D \, , \quad [D,K]=K\,,
\end{equation}

\begin{equation}\label{azcluke6}
[M_{ij},A_k]=\delta_{jk} A_i - \delta_{ik} A_j \,,
\end{equation}
where the subalgebra $A_i=(P_i,B_i,F_i)$ describes the generators of the  maximal Abelian subgroup. The algebra of  rotations $\emph{o}(d)$ is described by the commutators

\begin{equation}\label{azcluke7}
[M_{ij}, M_{k l}]= \delta_{ik} \,M_{j l} - \delta_{i l}\, M_{jk} +
\delta_{j l}\, M_{ik} - \delta_{jk} M_{i l}\,,
\end{equation}

\begin{equation}\label{azcluke8}
[M_{ij}, {\cal A}]=0\,, \qquad  {\cal A}= (H,D,K)\,.
\end{equation}
The generators ($P_i,B_i, H, M_{i j}$) define the $D$-dimensional Galilean algebra, where $B_i$ are the Galilean boosts and $H$, the non-relativistic energy operator, generates the Galilean time translations. We see that one can treat the Galilean conformal algebra as the
 result of adding the generators $F_i$ of constant accelerations  to the Schr\"{o}dinger algebra
  \cite{azcluk10}. We note that the subalgebra~(\ref{azcluke5}) is the one-dimensional conformal algebra $\emph{o}(2,1)$.

The aim of this paper is to introduce the supersymmetrization  of Galilean conformal symmetry. There have been several proposals 
to derive a supersymmetric Galilei algebra \cite{azcluk12}--\cite{azcluk16}, but among the conformal generalizations only the Schr\"{o}dinger symetry has been enlarged to a superSchr\"{o}dinger symmetry \cite{azcluk17}--\cite{azcluk20}. 
 These supersymmetric  Schr\"{o}dinger algebras have  not been obtained by performing the non-relativistic limit $c \to \infty$ of  relativistic superconformal algebra $\emph{su}(2,2|N)$, but rather  by considering a suitable projection of the relativistic odd generators (see e.g. \cite{azcluk19,azcluk20}). In this paper we shall use the natural procedure of the  non-relativistic contraction $c\to \infty$ limit  applied to the $N$-extended Wess-Zumino superconformal algebra $\emph{su}(2,2|2k)$ ($k=1,2,\ldots$). For odd $N$ (in particular for $N=1$) the method of non-relativistic contraction presented in this paper does not work.

The paper is organized as follows. In Sect.~2 we  describe the superconformal algebra $\emph{su}(2,2|N)$, which we shall rewrite further for $N=2k$ 
 by using  suitably projected supercharges. In Sect.~3 we  consider  our non-relativistic contraction limit $c \to \infty$, which leads to the Galilean superconformal algebra. From the relativistic superconformal internal sector $\emph{u}(2k)$ we obtain the  non-relativistic superconformal internal  symmetries $\emph{usp}(2k)\simeq \emph{sp}(k;H)$ $(k(2k+1)$ generators) and a set of $k(2k-1)$ Abelian central charges.

Finally in Sect.~4 we present an outlook.

\section{$N$-extended $D=4$ superconformal algebra $\emph{su}(2,2|N)$}

We  describe the $\emph{su}(2,2|N)$ superalgebra of antiHermitean generators\footnote{In the special $N=4$ case the axial charge  becomes central, and the $N=4$ superconformal algebra is often denoted as $\emph{psu}(2,2|4)$} 
 using the $D=4$ Majorana representation for the $4\times 4$ real Dirac matrices $\gamma_\mu \equiv (\gamma_\mu)_{\alpha}^{\ \beta}$ ($\{\gamma_\mu, \gamma_\nu \}=
 2\eta_{\mu\nu}$;  $\eta_{\mu\nu}= \hbox{diag}(-1,1,1,1)$) satisfying the properties
 
 \begin{equation}\label{azcluke10}
 \gamma_i = \gamma_i^T \,, \quad C=\gamma_0 = -\gamma_0 ^T  \,,\qquad
 \gamma_5= \gamma_0 \gamma_1 \gamma_2 \gamma_3 = - \gamma_5^T\,,
 \end{equation}
 where $C\equiv C_{\alpha\beta}$ describes the symplectic metric in the space of real Majorana spinors.

 a) \emph{bosonic sector} ($M_{\mu\nu}, P_\rho, D, K_\mu, T^{ab} \in \emph{u}(N)$)
 
 \begin{eqnarray}\label{azcluke11}
 &&[M_{\mu\nu}, M_{\rho\tau}] = \eta_{\mu \tau} M_{\nu \rho} 
 - \eta_{\mu \rho} M_{\nu \tau} + \eta_{\nu \rho} M_{\mu \tau}
 - \eta_{\nu \tau} M_{\mu\rho}\,,
 \nonumber \\[5pt]
 &&
 [M_{\mu\nu}, P_{\rho}] = \eta_{\mu \rho} P_{\nu} - \eta_{\nu\rho} P_\mu \,,
   \nonumber \\[5pt]
 &&
 [M_{\mu\nu}, K_{\rho}] = \eta_{\mu \rho} K_{\nu} - \eta_{\nu\rho} K_\mu \,,
  \nonumber \\[5pt]
 &&
 [P_\mu, P_\nu ] = [K_\mu, K_\nu ] = 0\,,
  \nonumber \\[5pt]
 &&
 [D, P _\mu ] = - P_\mu \,, \quad [D,K_\mu ] = K_\mu \,,
 \quad [D, M_{\mu\nu}]= 0\,,
  \nonumber \\[5pt]
 &&
 [P_\mu, K_\nu ] = 2 (\eta_{\mu \nu} D - M_{\mu\nu})\,.
 \end{eqnarray}
The internal superconformal symmetries (sometimes called $R$-symmetries) are    given  by the $\emph{u}(N)$ generators ($T^{ab}_{S}, i T^{ab}_{A}$)
where $T^{ab}_{S}=T^{ba}_{S}$ and
 $T^{ab}_{A}=-T^{ba}_{A}$; besides  $T^{ab}_{S} \in \frac{\emph{su}(N)}{\emph{o}(N)}$
 and 
 $T^{ab}_{A} \in \emph{o}(N)$. The internal $\emph{u}(N)$ Lie algebra can be described by the generators $T^{ab}_{S}$, $T^{ab}_{A}$ as follows
 
 \begin{eqnarray}\label{azcluke12}
 [T^{ab}_{S}, T^{cd}_{S}]
 &=&
 \delta^{bc}T^{ad}_{A}
 + \delta^{ac} T^{bd}_{A} 
 + \delta^{ad} T^{bc}_{A}
 + \delta^{bd} T^{ac}_{A}\,,
 \cr\cr
 [T^{ab}_{S}, T^{cd}_{A}]
 &=&
 \delta^{ad}T^{bc}_{S}
 + \delta^{bd} T^{ac}_{S} 
 - \delta^{ac} T^{bd}_{S}
 - \delta^{bc} T^{ad}_{S}\,,
 \cr\cr
 [T^{ab}_{A}, T^{cd}_{A}]
 &=&
 \delta^{bc}T^{ad}_{A}
 - \delta^{ac} T^{bd}_{A} 
 + \delta^{ad} T^{bc}_{A}
 - \delta^{ad} T^{bc}_{A}\, ,
 \end{eqnarray}
where we assume that $(T^{ab}_{S,A})^{\dagger}= - T^{ab}_{S,A}$.
One can introduce the $N\times N$ matrix representation of the algebra (\ref{azcluke12}) ($\tau^{ab}_{S}, \tau^{ab}_{A}$) which can be deduced from a $N\times N$ generalization of the Pauli matrices $\sigma_i$, supplemented by the unit matrix. For $N=2$ we get

\begin{equation}\label{azcluke13}
\tau^{ab}_{S} = (1^{ab}, \sigma^{ab}_1, \sigma^{ab}_3)\,,
\qquad
 \tau^{ab}_{A} = \varepsilon^{ab} = - i \sigma^{ab}_2\,.
 \end{equation}
 
 b) \emph{fermionic sector} ($Q^{a}_{\alpha}, S^{a}_{\alpha}$;  $ \, \alpha, \beta = 1 \ldots 4,\,  a,b=1, \ldots N$)

 \begin{subeqnarray}\label{azcluke14}
 \slabel{azcluke14a}
 \{Q^{a}_{\alpha}, Q^{b}_{\beta} \} & = & 2 \delta^{ab}(\gamma^\mu C)_{\alpha \beta} P_\mu\,,
 \\[5pt]
 \slabel{azcluke14b}
 \{S^{a}_{\alpha}, S^{b}_{\beta} \} & = & - 2 \delta^{ab}(\gamma^\mu C)_{\alpha \beta} K_\mu\,,
 \\[5pt]
 \slabel{azcluke14c}
 \{Q^{a}_{\alpha}, S^{b}_{\beta} \} & =& 
  \delta^{ab} [ 2C_{\alpha\beta} D - (\sigma^{\mu\nu} C)_{\alpha\beta }
 M_ {\mu\nu}
 - 4 (\gamma_5 C)_{\alpha \beta} A] 
 \nonumber \\[5pt]
 &&
 + 2C_ {\alpha\beta} T^{ab}_{A} + 2(\gamma_5 C)_{\alpha\beta} T^{ab}_{S}\,,
 \end{subeqnarray}

 \noindent
 where $(\gamma^\mu C)_{\alpha\beta} = (\gamma^\mu)_{\alpha}^{\ \gamma} C_{\gamma\beta}$ and the axial charge generator $A$ corresponds to the trace of $T^{ab}_{S}$ (in formula~(\ref{azcluke14c}) and further below we assume that $tr\, T_S \equiv T^{aa}_S =0$).
 
 c) \emph{mixed bosonic-fermionic sector (covariance relations)}

 \begin{eqnarray}\label{azcluke15}
 &
 [P_\mu , Q^{a}_{\alpha} ] = 0\, ,  \quad 
  [P_\mu, S^{a}_\alpha ] = - (\gamma_\mu)_{\alpha}^{\ \beta} Q^{a}_{\beta} \, ,&
 \nonumber \\[5pt]
 &
 [K_\mu, Q^{a}_\alpha ] = - (\gamma_\mu)_{\alpha}^{\ \beta} S^{a}_{\beta}\, ,
 \qquad 
 [K_\mu , S^{a}_{\alpha} ] = 0\,,
 &
 \nonumber \\[5pt]
 &
 [M_{\mu\nu} , Q^{a}_{\alpha}] = - \half
 (\sigma_{\mu\nu})_{\alpha}^{\ \beta} \,Q^{a}_{\beta}\, ,
 \qquad 
 [M_{\mu\nu}, S^{a}_{\alpha}] = - \half
 (\sigma_{\mu\nu})_{\alpha}^{\ \beta} \, S^{a}_{\beta}\, ,
 &
 \nonumber \\[5pt]
 &
 [D, Q^{a}_{\alpha}] = - \half Q^{a} _{\alpha}\,, 
  \qquad 
 [D, S^{a}_{\alpha}] =  \half \, S^{a}_{\alpha}\,,
 &
 \nonumber \\[5pt]
 &
 [A, Q^{a}_{\alpha}] =- \quart (1-\frac{4}{N}) (\gamma^5)_{\alpha}^{\ \beta} \, Q^{a}_{\beta}\,,
 &
 \nonumber \\[5pt]
 &
 [A, S^{a}_{\alpha}]= \quart(1- \frac{4}{N})
 (\gamma^5)_{\alpha}^{\ \beta} \, S^{a}_{\beta}\,, &
 \end{eqnarray}
 
 \begin{eqnarray}\label{azcluke16}
 [T^{ab}_{S}, Q^{C}_{\alpha}] 
 & = &
 (\gamma_5)_{\alpha}^{\ \beta} (\tau^{ab}_{S})^{cd} Q^{d}_{\beta}\,,
\nonumber \\[5pt]
 [T^{ab}_{S}, S^{C}_{\alpha}] 
 & = &
 -(\gamma_5)_{\alpha}^{\ \beta} (\tau^{ab} _{S})^{cd} S^{d}_{\beta}\,,
 \nonumber \\[5pt]
  [T^{ab}_{A}, Q^{C}_{\alpha}] 
 & = &
  (\tau^{ab} _{A})^{cd} Q^{d}_{\alpha}\,,
 \nonumber \\[5pt]
  [T^{ab}_{A}, S^{C}_{\alpha}] 
 & = &
  (\tau^{ab} _{A})^{cd} S^{d}_{\alpha}\,.
\end{eqnarray}

Further we introduce the following matrix projector operators

\begin{equation}\label{azcluke17}
(P_{\pm})^{ab}_{\alpha\beta} = \half (\delta_{\alpha\beta}\,
\delta^{ab} \pm C_{\alpha\beta}\,\Omega^{ab} )\,,
\end{equation}
where

\begin{equation}\label{azcluke18}
\Omega^{ab} = - \Omega^{ba} \, , \qquad
\Omega^2 = -1 \,.
\end{equation}
The definition (\ref{azcluke17}) leads to the following projection operator properties

\begin{equation}\label{azcluke19}
(P_{\pm})^{ab}_{\alpha\beta}
(P_{\pm})^{bc}_{\beta\gamma} =
(P_{\pm})^{ac}_{\alpha\gamma}\,,
\qquad
(P_{\pm})^{ab}_{\alpha\beta}(P_{\mp})^{bc}_{\beta\gamma}=0\,.
\end{equation}
We define the projected supercharges as follows

\begin{equation}\label{azcluke20}
Q^{a}_{\pm \alpha} = (P_{\pm})^{ab}_{\alpha\beta} \, Q^{b}_{\beta}\,,
\qquad 
S^{a}_{\pm \alpha} = (P_{\pm})^{ab}_{\alpha\beta} \, S^{b}_{\beta}\,.
\end{equation}
Using the symmetry property

 \begin{equation}\label{azcluke21}
 (P^{T}_{\pm})^{ab}_{\alpha\beta}
 \equiv (P_{\pm})^{ba}_{\beta\alpha} = (P_{\pm})^{ab}_{\alpha\beta}\,,
 \end{equation}
 one gets from (\ref{azcluke14a}--\ref{azcluke14c})
 
 \begin{eqnarray}\label{azcluke22}
 \{Q^{a}_{\pm \alpha}, Q^{b}_{\pm \beta}\} = 
 2 (P_{\pm})^{ab}_{\alpha\beta} \, P_0 \,,
 \nonumber \\[5pt]
  \{Q^{a}_{+\alpha}, Q^{b}_{- \beta}\} = 
 2 (P_{+})^{ab}_{\alpha\gamma} (\gamma_i C)_{\gamma\beta}\, P_i \,,
 \end{eqnarray}
 
 \begin{eqnarray}\label{azcluke23}
 \{S^{a}_{\pm \alpha}, S^{b}_{\pm \beta}\} = 
 -2 (P_{\pm})^{ab}_{\alpha\beta} \, K_0 \,,
 \nonumber \\[5pt]
  \{S^{a}_{+\alpha}, S^{b}_{- \beta}\} = 
 - 2 (P_{+})^{ab}_{\alpha\gamma} (\gamma_i C)_{\gamma\beta}\, K_i \,.
 \end{eqnarray}

 In order to rewrite the relation (\ref{azcluke14c}) in terms of the projected supercharges (\ref{azcluke20}) we split the generators $T^{ab}_{S}, T^{ab}_{A}$ into  the following four sectors
  
  \begin{eqnarray}\label{azcluke24}
  T^{ab}_{S} = T^{ab}_{S+} + T^{ab}_{S-}\, , \qquad
  T^{ab}_{A} = T^{ab}_{A+} + T^{ab}_{A-}\ \,,
 \end{eqnarray}
 where
   
   \begin{eqnarray}\label{azcluke25}
   [\, \Omega, T_{S+\atop A+} ]=0\,,
   \qquad
   \{\, \Omega, T_{S-\atop A-} \} =0\,.
   \end{eqnarray}
 Then, we obtain

  \begin{subeqnarray}\label{azcluke26}
  \slabel{azcluke26a}
 \{Q^{a}_{\pm \alpha}, S^{b}_{\pm \beta} \} &=&
 (P_{\pm})^{ac}_ {\alpha\gamma}
 [\delta^{cb} \{-(\sigma^{ij} C)_{\gamma\beta} \, M_{ij}
 \nonumber \\[5pt]
 && + 2 C_{\gamma\beta} D\} + 2 C_{\gamma\beta}
 T^{ab}_{A+} + 2 (\gamma_5 C)_{\gamma\beta} T^{ab}_{S-}]\,,
 \\ \cr
  \slabel{azcluke26b}
  \{Q^{a}_{\pm \alpha}, S^{b}_{\mp \beta} \} &=&
 (P_{\pm})^{ac}_ {\alpha\gamma}
 [\delta^{cb} \{-(\sigma^{0j} C)_{\gamma\beta} \, M_{0j}
 \nonumber \\[5pt]
 && 
 - 4 (\gamma_5 C)_{\gamma\beta} A\} + 2 C_{\gamma\beta}
 T^{ab}_{A-} + 2 (\gamma_5 C)_{\gamma\beta} T^{ab}_{S+}]\,.
  \end{subeqnarray}
 
 The relations~(\ref{azcluke25}) can be expressed also in the following way

  \begin{subeqnarray}\label{azcluke27}
  \slabel{azcluke27a}
  \Omega \cdot \mathbb{H} = - \mathbb{H}^{T} \cdot \Omega \, ,
  \qquad
  \mathbb{H} = (T_{S-}, T_{A+})\,,
  \\ \cr
  \slabel{azcluke27b}
  \Omega \cdot \mathbb{K} = \mathbb{K}^{T} \cdot \Omega \, ,
  \qquad
  \mathbb{K} = (T_{S+}, T_{A-})\,.
  \end{subeqnarray}
  Recalling that the $N\times N$  ($N=2k$) matrix representation of the generators of $\emph{u}(2k)$ when it is additionally constrained by the symplectic condition~(\ref{azcluke27a}) define the matrix subalgebra $\emph{usp}(2k)$ \cite{azcluk21},  we obtain

  \begin{equation}\label{azcluke28}
  \mathbb{H} = \emph{usp}(2k)\,, \qquad \quad  \mathbb{K} = \frac{\emph{u}(N)}{\emph{usp}(2k)}\,.
  \end{equation}
  Indeed using the table of $[A,B]$ commutators, obtained from the relations~(\ref{azcluke12}) and~(\ref{azcluke25})
 \begin{center}
 \begin{table}[h]
 \begin{center}
 \begin{tabular}{|l||l|l|l|l|}
 \hline
 $\quad \ A$ &&&&
 \\[-10pt]
 $ \quad \setminus$
  & $T_{S+}$ & $T_{S-}$ & $T_{A+}$ & $T_{A-}$
  \\[-10pt]
  $B$ &&&&
 \\
 \hline\hline
 $T_{S+}$ & $T_{A+}$  & $T_{A-}$ & $T_{S+}$ & $T_{S-}$
 \\ \hline
 $T_{S-}$ & $T_{A-}$  & $T_{A+}$ & $T_{S-}$ & $T_{S+}$
 \\ \hline
 $T_{A+}$ & $T_{S+}$  & $T_{S-}$ & $T_{A+}$ & $T_{A-}$
 \\ \hline
 $T_{A-}$ & $T_{S-}$  & $T_{S+}$ & $T_{A-}$ & $T_{A+}$
 \\ \hline
 \end{tabular}
 \end{center}
\caption{Fourfold split of $\emph{u}(N)$ algebra}
\end{table}
\end{center}
one can check the relations

\begin{equation}\label{azcluke29}
[\mathbb{H}, \mathbb{H} ] \subset \mathbb{H}\,,\qquad
[\mathbb{H}, \mathbb{K} ] \subset \mathbb{K} \,, \qquad
[\mathbb{K}, \mathbb{K}] \subset \mathbb{H}\,,
\end{equation}

\noindent
that describe the symmetric Riemannian space $(\mathbb{H}, \mathbb{K})$. The subalgebra $\mathbb{H}$ has $k(2k+1)$ generators, and the coset $\mathbb{K}$ contains $k(2k-1)$ generators. For low values of $k$ one gets:
 \begin{center}
 \begin{table}[h]
 \begin{center}
 \begin{tabular}{|l|c|c|c|}
 \hline
 $\quad \ \ k$ &&&
 \\[-12pt]
 $ \quad \setminus$
 & 1 & 2 & 4
 \\
 \hline
 $\mathbb{H}$ & $\emph{usp}(2)$ $\simeq \emph{su}(2)$ & $\emph{usp}(4)$ $\simeq \emph{o}(5)$ 
  & $\emph{usp}(8)$ 
\\
\hbox{dim} $\mathbb{H}$ & 3 & 10 & 36
\\
\hline
$\mathbb{K}$ & $\frac{\emph{u}(2)}{\emph{su}(2)}=\emph{u}(1)$ & $\frac{\emph{u}(4)}{\emph{usp}(4)} \simeq \emph{u}(1) \times \frac{\emph{o}(6)}{\emph{o}(5)}$ & $\frac{\emph{u}(8)}{\emph{usp}(8)}$
\\
\hbox{dim} ${\mathbb{K}}$ & 1 & 6 & 24
\\
\hline
\end{tabular}
\end{center}
\caption{Lower-dimensional symplectic cosets of $\emph{u}(2k)$}
\end{table}
\end{center}
Using the relations~(\ref{azcluke15}) and the projected supercharges~(\ref{azcluke20}) one gets the following set of non-zero commutators for the  mixed bosonic-fermionic sector

\begin{eqnarray}\label{azcluke30}
[P_i, S^{a}_{\pm \alpha}] 
& = &
-(\gamma_i)_{\alpha}^{\ \beta} \, Q^{a}_{\mp \beta} \,,
\nonumber\\[5pt]
[P_0, S^{a}_{\pm \alpha}] 
& = &
 Q^{a}_{\pm \alpha} \,,
 \nonumber\\[5pt]
 [K_i, Q^{a}_{\pm \alpha}] 
& = &
-(\gamma_i)_{\alpha}^{\ \beta} \, S^{a}_{\mp \beta} \,,
\nonumber\\[5pt]
[K_0, Q^{a}_{\pm \alpha}] 
& = &
S^{a}_{\pm \alpha} \,,
\nonumber\\[5pt]
[M_{ij}, Q^{a}_{\pm \alpha}] 
& = &
-\half (\sigma_{ij})_{\alpha}^{\ \beta} \, Q^{a}_{\pm \beta} \,,
\nonumber\\[5pt]
 [M_{0i}, Q^{a}_{\pm \alpha}] 
& = &
- \half (\sigma_{0i})_{\alpha}^{\ \beta} \, Q^{a}_{\mp \beta} \,,
\nonumber\\[5pt]
[M_{ij}, S^{a}_{\pm \alpha}] 
& = &
- \half(\sigma_{ij})_{\alpha}^{\ \beta} \, S^{a}_{\pm \beta} \,,
\nonumber\\[5pt]
[M_{0i}, S^{a}_{\pm \alpha}] 
& = &
- \half(\sigma_{0i})_{\alpha}^{\ \beta} \, S^{a}_{\mp \beta} \,,
\nonumber\\[5pt]
[D, Q^{a}_{\pm \alpha}] 
& = &
- \half \, Q^{a}_{\pm \alpha} \,, 
\qquad [D, S^{a}_{\pm \alpha}]= \half S^{a}_{\pm \alpha} \,,
\nonumber\\[5pt]
[A, Q^{a}_{\pm \alpha} ]
& = &
 - \quart
(1-\frac{4}{N})(\gamma^5)_{\alpha}^{\ \beta} \, Q^{a}_{\mp \beta}\,,
\nonumber\\[5pt]
[A, S^{a}_{\pm \beta} ]
& = &
  \quart
(1-\frac{4}{N})(\gamma^5)_{\alpha}^{\ \beta} \, S^{a}_{\mp \beta}\,.
\end{eqnarray}
 The projected supercharges~(\ref{azcluke20})
 permit to rewrite the relations~(\ref{azcluke16}) as follows:

\begin{eqnarray}\label{azcluke31}
[T^{ab}_{S+}, Q^{c}_{\alpha \pm}]
& = &
(\gamma_5)_{\alpha}^{\ \beta}\, (\tau^{ab}_{S+})^{cd}\, Q^{d}_{\beta\mp}\,,
\nonumber\\[4pt]
[T^{ab}_{S-}, Q^{c}_{\alpha \pm}]
& = &
(\gamma_5)_{\alpha}^{\ \beta}\, (\tau^{ab}_{S-})^{cd}\, Q^{d}_{\beta\pm}\,,
\nonumber\\[4pt]
[T^{ab}_{S+}, S^{c}_{\alpha \pm}]
& = &
-(\gamma_5)_{\alpha}^{\ \beta}\, (\tau^{ab}_{S+})^{cd}\, S^{d}_{\beta\mp}\,,
\nonumber\\[4pt]
[T^{ab}_{S-}, S^{c}_{\alpha \pm}]
& = &
-(\gamma_5)_{\alpha}^{\ \beta}\, (\tau^{ab}_{S-})^{cd}
 S^{d}_{\alpha\pm}\,,
\nonumber\\[4pt]
[T^{ab}_{A+}, Q^{c}_{\alpha \pm}]
& = &
 (\tau^{ab}_{A+})^{cd}\, Q^{d}_{\alpha\pm}\,,
\nonumber\\[4pt]
[T^{ab}_{A-}, Q^{c}_{\alpha \pm}]
& = &
 (\tau^{ab}_{A-})^{cd}\, Q^{d}_{\alpha\mp}\,,
\nonumber\\[4pt]
[T^{ab}_{A+}, S^{c}_{\alpha \pm}]
& = &
 (\tau^{ab}_{A+})^{cd}\, S^{d}_{\alpha\pm}\,,
\nonumber\\[4pt]
[T^{ab}_{A-}, S^{c}_{\alpha \pm}]
& = &
 (\tau^{ab}_{A-})^{cd}\, S^{d}_{\alpha\mp}\,.
\end{eqnarray}

\noindent
The action of the generators  $(T_{S-}, T_{A+})\in \mathbb{H}$ produces  linear transformations on the projected supercharges, and  the generators $(T_{S+}, T_{A-})\in \mathbb{K}$ transform  $(Q^{a}_{\alpha \pm}, S^{a}_{\alpha \pm})$ into the complementary projections $(Q^{a}_{\alpha \mp}, S^{a}_{\alpha \mp})$.

\section{Galilean $(N=2k)$-extended superconformal algebra as the non-relativistic contraction of  $\emph{su}(2,2|2k)$}

In this Section we  perform the contraction of  the $N$-extended 
($N=2k$) $D=4$ relativistic conformal superalgebras in a way that provides the supersymmetrization of Galilean conformal algebra, (see Eqs.~(\ref{azcluke2}--\ref{azcluke8})). To achieve this contraction with finite contraction limits we supplement the bosonic rescalings~(\ref{azcluke1}) by the following ones

\begin{subeqnarray}\label{azcluke32}
\slabel{azcluke32a}
Q^{a}_{\alpha +} = \frac{1}{\sqrt{c}} \, \tilde{Q}^{a}_{\alpha +}\,,
\qquad Q^{a}_{\alpha -} = \sqrt{c} \, \tilde{Q}^{a}_{\alpha -}\,,
\\
\slabel{azcluke32b}
S^{a}_{\alpha +} = {\sqrt{c}} \, \tilde{S}^{a}_{\alpha +}\,,
\qquad S^{a}_{\alpha -} = ({c})^{3/2} \, \tilde{S}^{a}_{\alpha -}\,,
\end{subeqnarray}
where the tilde denotes the Galilean algebra supercharges\footnote{The relative factor $c$ between the two projections $Q^{a}_{\pm \alpha}$, $S^{a}_{\pm \alpha}$ can be justified if we observe that
  the projector $P_{\pm}=\half (1\pm \gamma^{0})$ separates the ``large'' and ``small'' components of a Dirac spinor, a fact that has been used before \cite{azcluk13,azcluk15}. The global factor $c$ between the $Q$ and the $S$ rescalings follows from dimensional considerations.}.
Further we introduce the rescaling (see~(\ref{azcluke27a}))

\begin{eqnarray}\label{azcluke33}
h^{ab} \equiv (T^{ab}_{S-}, T^{ab}_{A+} ) =
(\tilde{T}^{ab}_{S-} , \tilde{T}^{ab}_{A+} )\,,
\nonumber \\[4pt]
k^{ab} \equiv (A,{T}^{ab}_{S+}, {T}^{ab}_{A-}) =
(c\tilde{A}, c \tilde{T}^{ab}_{S+}, c\tilde{T}^{ab}_{A-})\,,
\end{eqnarray}
where $h^{ab} \in \mathbb{H}$, $k^{ab} \in \mathbb{K}$,
 and $\tilde{T}^{ab}_S, \tilde{T}^{ab}_{A}, \tilde{A}$ denote the Galilean internal symmetry generators or, more briefly,
\begin{equation}\label{azcluke34}
(\mathbb{H}, \mathbb{K}) = (\tilde{\mathbb{H}}, c\tilde{\mathbb{K}})\,.
\end{equation}
{}From~(\ref{azcluke29}) we get
\begin{equation}\label{azcluke35}
[\tilde{\mathbb{H}}, \tilde{\mathbb{H}}] \subset \tilde{\mathbb{H}}\,,
\qquad
 [\tilde{\mathbb{H}}, \tilde{\mathbb{K}}] \subset \tilde{\mathbb{K}}\,,
\end{equation}
and
\begin{equation}\label{azcluke36}
[\tilde{\mathbb{K}}, \tilde{\mathbb{K}}] \subset \frac{1}{c^2} \tilde{\mathbb{H}}
\mathop{\longrightarrow}\limits_{ c \to \infty } 0
\end{equation}
i.e. the generators $\tilde{k}_{ab} \in \tilde{\mathbb{K}}$ become Abelian.
As a result of the contraction we see that  the Galilean internal symmetry algebra
\begin{equation}\label{azcluke37}
\tilde{\mathbb{T}} = \tilde{\mathbb{K}}
 \supsetplus
 \tilde{\mathbb{H}} 
 = T^{k(2k-1)} 
 \supsetplus
  \emph{usp}(2k)  
\end{equation}
is non-semisimple.

We  perform now the contraction $c\to \infty$ of the relations~(\ref{azcluke22}--\ref{azcluke23}), (\ref{azcluke26a}--\ref{azcluke26b}) and (\ref{azcluke30}--\ref{azcluke31}). The algebraic relations that describe the $2k$-extended Galilean superconformal algebra are:

a) \emph{Fermionic sector}

\begin{subeqnarray}\label{azcluke38}
\slabel{azcluke38a}
\{ \tilde{Q}^{a}_{+\alpha}, \tilde{Q}^{b}_{+ \beta}\}
&= &
2(P_{\pm})^{ab}_{\alpha\beta} \, H \,, 
\qquad 
\{\tilde{Q}^a_{- \alpha}, \tilde{Q}^b_{-\beta}\} = 0 \,,
\nonumber\\[2pt]
\{ \tilde{Q}_{+ \alpha}, \tilde{Q}_{- \beta} \}
& = &
 2(P_{+})^{ab}_{\alpha\gamma}(\gamma_i C)_{\gamma\beta} \, P_i\,,
 \\[4pt]
\slabel{azcluke38b}
\{ \tilde{S}^{a}_{+ \alpha}, \tilde{S}^{b}_{+ \beta} \}
& = &
-2(P_{\pm})^{ab}_{\alpha\beta} \, K\,,
\qquad 
\{\tilde{S}^a_{- \alpha}, \tilde{S}^b_{-\beta}\} = 0 \,,
\nonumber \\[2pt]
\{ \tilde{S}^{a}_{+ \alpha}, \tilde{S}^{b}_{- \beta} \}
& = &
-2(P_{+})^{ab}_{\alpha\gamma} (\gamma_i C)_{\gamma\beta} \, F_i\,,
\\[4pt]
\slabel{azcluke38c}
\{ \tilde{Q}^{a}_{+ \alpha}, \tilde{S}^{b}_{+ \beta} \}
& =&
(P_{+})^{ac}_{\alpha\gamma} 
[\delta^{cb}\{-(\sigma^{ij} C)_{\gamma\beta}\, M_{ij}
\nonumber \\[2pt]
&& +2 C_{\alpha\beta}D\} + 2 C_{\gamma\beta}
\tilde{T}^{cb}_{A+} 
+(2\gamma_5 C)_{\gamma\beta} \tilde{T}^{cb}_{S-}]\,,
 \\[4pt]
\{\tilde{Q}^a_{-\alpha} , \tilde{S}^b_{-\beta}\}& = & 0\,,
\nonumber \\[4pt]
\slabel{azcluke38d}
\{ \tilde{Q}^{a}_{\pm \alpha}, \tilde{S}^{b}_{\mp \beta} \}
& = &
(P_{\pm})^{ac}_{\alpha\gamma} 
[-\delta^{cb}\{(\sigma^{0j} C)_{\gamma\beta}\, B_{j}
\nonumber \\[2pt]
&& - 4 (\gamma_5 C)_{\gamma\beta}\tilde{A}\}
+2 C_{\gamma\beta} \tilde{T}^{cb}_{A-}
+2(\gamma_5 C)_{\gamma\beta} \tilde{T}^{cb}_{S+}]\,,
\end{subeqnarray}
where in Eq.~(\ref{azcluke38c})  only remain the  generators which belong to the subalgebra $\tilde{\mathbb{H}}$, and in Eq.~(\ref{azcluke38d}) appear  only the central charges obtained from the generators belonging to $\mathbb{K}$ (see Eqs.~(\ref{azcluke33}) and~(\ref{azcluke36})).

b) \emph{Mixed bosonic-fermionic sector }

After the rescalings (\ref{azcluke1}), (\ref{azcluke32a}) 
 we obtain from (\ref{azcluke30})
in the limit $c\to \infty$

\begin{equation}\label{azcluke39}
\begin{array}{llll}
[P_i, \tilde{S}^{a}_{+\alpha}] 
&=&
- (\gamma_i)_{\alpha}^{\ \beta} \tilde{Q}^{a}_{-\beta}\, ,
\quad 
&[P_i, \tilde{S}^{a}_{-\alpha}] 
=
0\,,
\\[5pt]
[H, \tilde{S}_{\pm \alpha} ]
& = &
\tilde{Q}^{a}_{\pm \alpha}\,,
 \\[5pt]
[F_i, \tilde{Q}^{a}_{-\alpha} ]
& = &
-(\gamma_i)_{\alpha}^{\ \beta} \tilde{S}^{a}_{+\beta}\, ,
\quad 
&[F_i, \tilde{Q}^{a}_{+\alpha} ]
 = 
0 \,,
 \\[5pt]
[K, \tilde{Q}^{a}_{\pm \alpha} ]
& = &
\tilde{S}^{a}_{\pm \alpha}\,,
 \\[5pt]
[M_{ij}, \tilde{Q}^{a}_{\pm\alpha}] 
&=&
- \half(\sigma_{ij})_{\alpha}^{\ \beta} {Q}^{a}_{\pm\beta}\,,
 \\[5pt]
[B_{i}, \tilde{Q}^{a}_{+\alpha}] 
&=&
- \half(\sigma_{0i})_{\alpha}^{\ \beta} \tilde{Q}^{a}_{-\beta}\,,
\\[5pt]
[B_{i}, \tilde{Q}^{a}_{-\alpha}] 
&=&
0\, ,
\quad 
&[M_{ij}, \tilde{S}^{a}_{\pm\alpha}] 
 = 
- \half(\sigma_{ij})_{\alpha}^{\ \beta} \tilde{S}^{a}_{\pm\beta}\, ,
 \\[5pt]
[B_{i}, \tilde{S}^{a}_{+\alpha}] 
& = &
0 \, ,
\quad 
&[B_{i}, \tilde{S}^{a}_{-\alpha}] 
 =
- \half(\sigma_{0i})_{\alpha}^{\ \beta} \tilde{S}^{a}_{+\beta}\,,
\\[5pt]
[D, \tilde{Q}^{a}_{\pm\alpha}] 
&=&
- \half \tilde{Q}^{a}_{\pm \alpha}\,,
\quad 
&[D, \tilde{S}^{a}_{\pm \alpha}] = \half \tilde{S}^{a}_{\pm \alpha}\,,
 \\[5pt]
[\tilde{A}, \tilde{Q}^{a}_{+\alpha}] 
&=&
- \quart (1- \frac{4}{N})(\gamma_5)_{\alpha}^{\ \beta} \tilde{Q}^{a}_{-\beta}\, ,
\quad
&[\tilde{A}, \tilde{Q}^{a}_{-\alpha} ] 
 = 
0\,,
\\[5pt]
[\tilde{A}, \tilde{S}^{a}_{+\alpha}] 
&=&
 \quart (1- \frac{4}{N})(\gamma_5)_{\alpha}^{\ \beta} \tilde{S}^{a}_{-\beta}\, ,
 \quad 
 &[\tilde{A}, \tilde{S}^{a}_{-\beta} ]
  = 
 0 \, .
\end{array}
\end{equation}

The  relations (\ref{azcluke31}) that give the transformation  properties under the internal symmetry generators have the following non-relativistic limits:

i) for the generators $\tilde{h}^{ab}= ( \tilde{T}_{S-}, \tilde{T}_{A +})$
\begin{eqnarray}\label{azcluke40}
[\tilde{T}^{ab}_{S-} , \tilde{Q}^{c}_{\alpha \pm} ]
& = &
- (\gamma_5)_{\alpha}^{\ \beta} (\tau^{ab}_{S-})^{cd} \, \tilde{Q}^{d}_{\beta \pm}\,,
\nonumber \\[2pt]
[\tilde{T}_{A+} , \tilde{Q}^{c}_{\alpha \pm} ]
& = &
 (\tau^{ab}_{A+})^{cd} \, \tilde{Q}^{d}_{\alpha \pm}\,,
\nonumber \\[2pt]
[\tilde{T}_{S-} , \tilde{S}_{\alpha \pm} ]
& = &
- (\gamma_5)_{\alpha}^{\ \beta} (\tau^{ab}_{S-})^{cd} \, \tilde{S}^{d}_{\beta \pm}\,,
\nonumber \\[2pt]
[\tilde{T}_{A+} , \tilde{S}_{\alpha \pm} ]
& = &
(\tau^{ab}_{A+})^{cd} \, \tilde{S}^{d}_{\alpha \pm}\,.
\end{eqnarray}

ii) for the generators $\tilde{k}^{ab} = (\tilde{T}_{S+}, \tilde{T}_{A-})$

\begin{equation}\label{azcluke41}
\begin{array}{llll}
[\tilde{T}^{ab}_{S+} , \tilde{Q}^{c}_{\alpha +} ]
 & = &
- (\gamma_5)_{\alpha}^{\ \beta} (\tau^{ab}_{S+})^{cd} \, \tilde{Q}^{d}_{\beta -}\,,
\quad \ \ \,
&
[\tilde{T}^{ab}_{S+} , \tilde{Q}^{c}_{\alpha -} ]
 = 
0\,,
 \\[4pt]
[\tilde{T}^{ab}_{A-} , \tilde{Q}^{c}_{\alpha +} ]
& = &
 (\tau^{ab}_{A-})^{cd} \, \tilde{Q}^{d}_{\alpha -}\, ,
\qquad \qquad \
&
[\tilde{T}^{ab}_{A-} , \tilde{Q}^{c}_{\alpha -} ]
 = 
0\,,
 \\[4pt]
[\tilde{T}^{ab}_{S+} , \tilde{S}^{c}_{\alpha +} ]
& = &
- (\gamma_5)_{\alpha}^{\ \beta} (\tau^{ab}_{S+})^{cd}  \tilde{S}^{d}_{\beta -}\,,
\qquad \ \   
&
[\tilde{T}^{ab}_{S+} , \tilde{S}^{c}_{\alpha -} ]
  = 
0\,,
\\[4pt]
[\tilde{T}^{ab}_{A-} , \tilde{S}^{c}_{\alpha +} ]
& = &
(\tau^{ab}_{A-})^{cd}  \tilde{S}^{d}_{\alpha -}\, ,
\qquad \qquad \ \,
&
[\tilde{T}^{ab}_{A-} , \tilde{S}^{c}_{\alpha -} ]
  =  
0\,.
\end{array}
\end{equation}

c) \emph{Bosonic sector of Galilean superconformal algebra}

The spacetime bosonic sector is described by the $D=4$ Galilean conformal algebra (see~(\ref{azcluke2})--(\ref{azcluke8}))
and the internal bosonic sector is the inhomogeneous $T^{k(2k-1)} 
\supsetplus
 \emph{usp}(2k)$ algebra~(\ref{azcluke37}). These two sectors commute in agreement with the Coleman-Mandula theorem \cite{azcluk22}.
 
 \section{Outlook}
 In this paper we have introduced  a contraction prescription that provides a  supersymmetrization of the Galilean conformal algebra, containing the non-relativistic conformal translations that describe constant accelerations. There are some further questions to be studied:
 
 i) The contraction presented in this paper which selects as distinguished translation generator the single Hamiltonian is well adapted to the description of point particle dynamics. If we move to  considering  $p$-brane dynamics 
 before performing the non-relativistic limit 
 we  should split 
  the relativistic translation generators into the ones that describe the shift of world volume coordinates and  those which correspond to the directions transversal to the $p$-brane \cite{azcluk23,azcluk24}. 
 There have been already some proposals 
   to describe the non-relativistic superconformal symmetries  
   in such a framework (see e.g. \cite{azcluk25,azcluk26}). In this paper the  rescaling is given by Eq.~(\ref{azcluke32a}), which is the way to obtain a
    standard   non-relativistic superconformal limit $c \to \infty$.
 
 ii) Other important question is the description of the 
 quantum-mechanical realizations of our Galilean superconformal algebra. In \cite{azcluk8} a realization of the Galilean conformal algebra  in $D=2+1$-dimensional non-relativistic phase space was proposed, but the corresponding action \cite{azcluk26} is defined by a  Lagrangian  with higher order time derivatives.
 It would be interesting to see using e.g. the techniques presented in \cite{azcluk27}  how the appearance of central charges in our scheme can  be used for the description of new classes of non-relativistic superconformal mechanics\footnote{Such    superconformal mechanics should be not confused  with the models providing  $N$-extended conformal supersymmetry on the world line, described by $\emph{osp}(N|2)$ superalgebra.}
 
 iii) An interesting question is to understand in a more fundamental way whether, 
  as follows from our method, there are no
    $N$-extended Galilean superconformal symmetries for $N$ odd (e.g. $N=1$).
    
    \subsection*{Note added:} The Galilean superconformal algebra was found independently by M.~Sakaguchi [29], where  a generalization of Galilean conformal and superconformal algebras for non-relativistic systems with finite space curvature described by Newton-Hooke geometries
    were also discussed. There, a rescaling of $P_i$ leaving $P_0$ unchanged was used, whereas in this paper we have supersymmetrically extended the standard nonrelativistic rescaling of $P_0$ (Eq.~(1)) leaving $P_i$ unchanged.
     The authors would like to thank M.~Sakaguchi for spotting some misprints in our paper.
 
 \subsection*{Acknowledgments}
 We would like to thank Sergey Fedoruk, Joaquim Gomis and Peter Horvathy  for discussions.
 J.A. wishes to thank the Spanish Ministry of Science for the research grant FIS2008-01980; 
  J.L. would like to acknowledge the financial support of Polish Ministry of Science and Higher Education grant NN202318534.


\begin{thebibliography}{99}

\bibitem{azcluk1} C. Duval, P. Horvathy, \emph{ ``Non-relativistic conformal symmetries and Newton-Cartan structures'' }, arXiv:0904.0531 [hep-th].
\bibitem{azcluk2} C.R. Hagen, Phys. Rev. \textbf{D5}, 377 (1972).
\bibitem{azcluk3} U. Niederer, Helv. Phys. Acta \textbf{45}, 802 (1972).
\bibitem{azcluk4} C. Burdet, M. Perrin, Lett. Nuovo Cim. \textbf{4}, 651 (1972).
\bibitem{azcluk5} M. Henkel, J. Stat. Phys. \textbf{75}, 1023 (1994).
\bibitem{azcluk6} Y. Nishida, D.T.~Son, Phys. Rev. \textbf{D76}, 086004 (2007).
\bibitem{azcluk7} A. Galajinsky, Phys. Rev. \textbf{D78}, 044014 (2008).
\bibitem{azcluk8} J. Lukierski, P. Stichel, W.J.~Zakrzewski, Phys. Lett. \textbf{B357}, 1 (2006).
\bibitem{azcluk9} A. Bagchi, R. Gopakumar, \emph{ ``Galilean Conformal Algebras and AdS/CFT'' }, arXiv:0902.1385 [hep-th].
\bibitem{azcluk10} J. Lukierski, PC.~Stichel, W.J.~Zakrzewski, Phys. Lett. \textbf{B650}, 203 (2007).
\bibitem{azcluk11} J.~Negro, M.A. del~Olmo, A. Rodr\'{\i}guez-Marco , J. Math. Phys. \textbf{38}, 3786 (1997); ibid. 3810 (1997).
\bibitem{azcluk12} R. Puzalowski, Acta Phys. Austr. \textbf{50}, 45 (1978).
\bibitem{azcluk13} J.A. de Azc\'{a}rraga, D. Ginestar, J. Math. Phys. \textbf{32}, 3500 (1991).
\bibitem{azcluk14} O. Bergmann, C.B.~Thorn, Phys. Rev. \textbf{D52}, 5980 (1995).
\bibitem{azcluk15} J. Gomis, K. Kamimura, P. Townsend, JHEP0411-051 (2004).
\bibitem{azcluk16} J. Lukierski, I. Pr\'{o}chnicka, P.~Stichel, W.J.~Zakrzewski, Phys. Lett. \textbf{B639}, 389 (2006).               
\bibitem{azcluk17} J.~Gauntlett, J. Gomis, P.~Townsend, Phys. Lett.
\textbf{B248}, 288 (1990).

\bibitem{azcluk18} C. Duval, P.A. Horvathy, J. Math. Phys. \textbf{35}, 2516 (1994).

\bibitem{azcluk19} M. Henkel, J. Unterberger, Nucl. Phys. \textbf{B746}, 155 (2006).

\bibitem{azcluk20} M. Sakaguchi, K. Yoshida, \emph{ ``SuperSchr\"{o}dinger in Super Conformal''}, arXiv:0805.2661   [hep-th]; ``More super Schr\"{o}dinger algebras from $psu(2,2|4)$, arXiv:0806.3612 [hep-th]

\bibitem{azcluk21} R. Gilmore, \emph{ Lie Groups, Lie      Algebras and Some of Their Applications}; Ed. J. Wiley and Sons, New York, 1974.

\bibitem{azcluk22} S. Coleman, J. Mandula, Phys. Rev. \textbf{150}, 1251 (1967).

\bibitem{azcluk23} J. Brugues, T. Curtright, J. Gomis, L.~Mezincescu, Phys. Lett. \textbf{B594}, 227 (2004).

\bibitem{azcluk24} J. Gomis, K. Kamimura, P. West, Class. Quant. Grav. \textbf{23}, 7369 (2006).

\bibitem{azcluk25} J. Gomis, J. Gomis, K.Kamimura, JHEP0512024 (2005).

\bibitem{azcluk26} M. Sakaguchi, K. Yoshida, JHEP0610:078~(2006).

\bibitem{azcluk27} P. Stichel, W.J.~Zakrzewski, Ann. Phys. 
\textbf{310}, 158 (2004).

\bibitem{azcluk28} J. Gomis, J. Lukierski, Phys. Lett. \textbf{B664}, 107 (2008).


\bibitem{azcluk29} M. Sakaguchi, \emph{``Super Galilean conformal algebra in AdS/CFT''}, arXiv:0905.0188 [hep-th].
\end{thebibliography}
\end{document}